\pdfoutput=1

\documentclass[11pt]{article}

\usepackage[preprint]{acl}

\usepackage{times}
\usepackage{latexsym}
\usepackage{multirow}
\usepackage{graphicx}
\usepackage{hyperref}
\usepackage{xcolor}

\usepackage[T1]{fontenc}

\usepackage[utf8]{inputenc}

\usepackage{microtype}

\usepackage{inconsolata}

\title{Automated Detection and Analysis of Data Practices Using A Real-World Corpus}

\author{Mukund Srinath\\
  Penn State University\\
  \texttt{mukund@psu.edu} \\\And 
  Pranav Venkit\\
  Penn State University\\
  \texttt{pranav.venkit@psu.edu} \\\And
  Maria Badillo\\
  The Future of Privacy Forum\\
  \texttt{mbadillo@fpf.org}
  \AND
  Florian Schaub\\
  University of Michigan\\
  \texttt{fschaub@umich.edu}\\\And
  C. Lee Giles\\
  Penn State University\\
  \texttt{clg20@psu.edu}\\\And
  Shomir Wilson\\
  Penn State University\\
  \texttt{shomir@psu.edu}
    }

\begin{document}
\maketitle
\begin{abstract}
Privacy policies are crucial for informing users about data practices, yet their length and complexity often deter users from reading them. In this paper, we propose an automated approach to identify and visualize data practices within privacy policies at different levels of detail. Leveraging crowd-sourced annotations from the ToS;DR platform, we experiment with various methods to match policy excerpts with predefined data practice descriptions. We further conduct a case study to evaluate our approach on a real-world policy, demonstrating its effectiveness in simplifying complex policies. Experiments show that our approach accurately matches data practice descriptions with policy excerpts, facilitating the presentation of simplified privacy information to users. 

\end{abstract}

\section{Introduction}

Do internet users care about their online privacy? While studies have shown that users care about their privacy online \citep{spiekermann2001privacy}, they are also willing to give away their personal information \citep{barnes2006privacy}. 
Studies have found that this is often because, \textit{privacy policies}, which are intended to inform users about what happens with their data online, are notoriously difficult to understand and time-consuming to read \cite{obar2018biggest}. 
An average user needs to spend \(\sim\)200 hours to read all the policies that they come across each year \cite{mcdonald2008cost}. Moreover, privacy policies are written at a college reading level \cite{ermakova2015readability}, use complicated legal jargon and are hard to comprehend \cite{meiselwitz2013readability}.

Some studies have attempted to ease the design of privacy policies \citep{schaub2015design} to make them more comprehensible to users. \citet{kelley2010standardizing} found that presenting privacy information visually in the form of a \textit{privacy nutrition label} significantly improved the ability of users to find and understand privacy information. While companies such as Apple and Google have incorporated a \textit{privacy label} as a way to systematically present users with concise summaries of an app’s data practices, they often fail to answer important privacy questions \citep{zhang2023privacy}. Moreover, most organizations do not provide access to easy-to-read labels. Given that there are several issues related to conducting privacy policy research \citep{mhaidli2023researchers}, scaling the creation of such labels using NLP approaches is a non-trivial task.

\begin{figure}[t]
    \centering
    \includegraphics[scale=0.45]{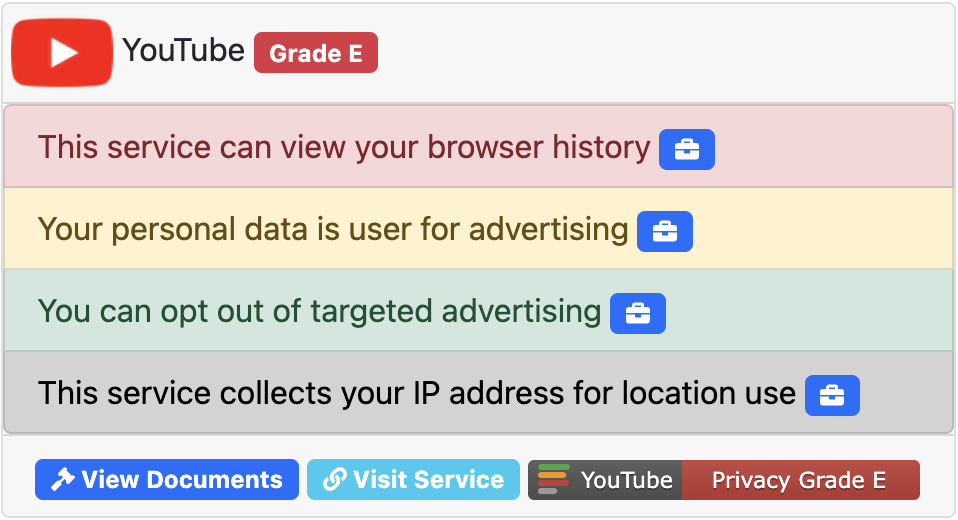}
    \caption{Snapshot of annotated ToS;DR \textit{cases} and assigned privacy letter grade for YouTube}
    \label{fig:tosdr_snapshot}
\end{figure}

We hence propose to automatically present users information regarding their online privacy using data practice ratings (blocker/good/bad/neutral) and descriptions derived from the Terms of Service; Didn't Read (ToS;DR) platform \cite{roy2012terms}. We make the following contributions:\\
$\bullet$ We manually analyze ToS;DR data descriptions, and cluster similar data practices together\footnote{Data and code included in the supplementary material}. \\
$\bullet$ We create an approach to automatically match excerpts from policies to easily understandable data practice descriptions. \\
$\bullet$ We design a privacy label to provide users with information regarding their data privacy at various levels of detail. 

\section{Related Work}

Various NLP methods have been employed to analyze privacy policies, yet none offer a holistic comprehension on data practices. Approaches centered around question answering \citep{ravichander2019question, ahmad2020policyqa},  summarization \citep{zimmeck2014privee, zaeem2018privacycheck} and classification \citep{wilson2016creation, nokhbeh2022privacycheck, tesfay2018privacyguide, 10.1145/3573128.3609342, sundareswara2021large} often lack comprehensive coverage of all data practices outlined in a policy \cite{nokhbeh2020privacycheck, nokhbeh2022privacycheck}. Moreover, these techniques place the onus on users to identify their specific information needs \cite{srinath2021privaseer, 10.1145/3573128.3604902, sundareswara2020privacy}. Users also face the challenge of investing substantial time in deciphering the nuances of each encountered policy \citep{meier2020shorter}. 

Previous studies categorized internet users into 5 categories based on varying levels of privacy concern and online privacy-related behaviors \citep{kumaraguru2005privacy, dupree2016privacy}. Given that each user category necessitates a different level of privacy information, there is a clear need for an approach that addresses this diversity. Our solution caters to this requirement.

\section{Methodology}

\subsection{Dataset}

Terms of Service; Didn't read (ToS;DR) aims to make understanding privacy policies easier by crowdsourcing annotations on them \cite{roy2012terms}. 
On the platform, internet users sign up to annotate policies by matching a policy excerpt to a pre-defined data practice (called a ``case'' on the platform). Once an annotator makes an excerpt-case match, a ToS;DR moderator either accepts the annotation or rejects it with comments. When a threshold percent of moderator approved \textit{case}-excerpt matches are annotated for a policy, it is given a letter grade. 

ToS;DR comprises 130 privacy policy-related \textit{cases}. A case on ToS;DR contains an average of  83 policy excerpts (min: 2; max: 83), each from a different service. Eaxt excerpt has an average length of 32.26 words/excerpt. Some \textit{cases} describe similar but contrasting data practices i.e. different approaches to handling the same data type. We deemed these as \textbf{contrasting cases}. Following is an example of two contrasting cases:\\
`\textit{Your personal data is not sold}' \\
`\textit{Your personal data is sold unless you opt out}' \\ 
Two authors manually evaluated the cases in ToS;DR and grouped \textit{contrasting cases} into 24 clusters (min: 2; max: 6 excerpts/cluster).  We note that \textit{contrasting cases} might often be a negation of one another, or might differ by describing a special circumstance of a general case, such as, \\
`\textit{You can delete your content from this service}'\\
`\textit{You cannot delete your content from this service}'\\
`\textit{You cannot delete your content, but it makes sense for this service}' \\
The remaining 67 \textit{cases} had no contrasting data practices. We deemed these as \textbf{standalone cases}.

Based on whether \textit{cases} preserve or erode user privacy, they are pre-labeled with one of the following rating categories: \textit{blocker}, \textit{bad}, \textit{neutral}, or \textit{good}. Of the 130 cases, 63 were rated good, 35 bad, 25 neutral, and 7 blockers. Figure \ref{fig:tosdr_snapshot} shows a snapshot from ToS;DR for the service \textit{YouTube} containing YouTube's assigned letter grade, selected cases, and their respective ratings (Blocker: Red; Bad: Yellow; Good: Green; Neutral: Grey). Ratings indicate respect towards user privacy, where `blockers' often indicate practices that are aggressively adversarial towards user privacy, while those rated `good' indicate privacy-preserving practices. `Bad' and `neutral' fall in the middle.

\subsection{Experiments}

\begin{table*}[ht]
\centering
\scalebox{1}{
\begin{tabular}{|c|c|c|c|c|c|c|c|c|c|}
\hline
\multirow{2}{*}{\textbf{Model}} & \multirow{2}{*}{\textbf{C}} &  \multicolumn{4}{|c|}{\textbf{Standalone Set}} & \multicolumn{4}{|c|}{\textbf{Contrasting Set}} \\
\cline{3-10}
 & & \textbf{P} & \textbf{R} & \textbf{F1} & \textbf{S} & \textbf{P} & \textbf{R} & \textbf{F1} & \textbf{S} \\ 
\hline
\multirow{2}{*}{\(RS\)} & 0 & \textbf{0.94} & 0.95 & \textbf{0.94} & 2091 & \textbf{0.75} & 0.37 & 0.50 & 557\\
                        & 1 & 0.95 & \textbf{0.94} & \textbf{0.94} & 2091 & 0.72 & 0.93 & 0.81 & 996\\
\hline
\multirow{2}{*}{\(CBS\)}& 0 & 0.89 & \textbf{0.95} & 0.92 & 2091 & 0.71 & \textbf{0.82} & \textbf{0.76} & 557\\
                        & 1 & \textbf{0.95} & 0.88 & 0.91 & 2091 & \textbf{0.95} & \textbf{0.91} & \textbf{0.93} & 996\\
\hline
\end{tabular}}
\caption{\label{held_out_eval} Performance of PrivBERT on random sampling and cluster-based sampling with a 1:3 sampling ratio (P: Precision, R: Recall, S: Support)}
\end{table*}

In this section, we describe our approach to automatically match excerpts from privacy policies to their respective ToS;DR \textit{cases}. Formally, given an excerpt from a privacy policy, \(S_j\), the task is to find the case(s) \(C_i\) that represent the data practice described in the excerpt \(S_j\). Here, \(C\) is the set of 130 ToS;DR \textit{cases}; S is the set of excerpts from a privacy policy. This problem can be framed as a multi-label classification task or a binary classification task. Preliminary experiments indicated that the multi-label classification was ineffective, due to a substantial imbalance in class distribution. Therefore, choose a binary classification technique to train a model to distinguish between matching and disparate case-excerpt pairs.
 
Annotated case-excerpt pairs from ToS;DR serve as positive training samples. To collect negative samples, we use two sampling approaches, random sampling (RS) and cluster-based sampling (CBS).\\
\textbf{Random Sampling (RS):} For each annotated ToS;DR \textit{case}-excerpt pair, we randomly sample excerpts matched with other \textit{case} to serve as a negative sample irrespective of it being a \textit{contrasting} or \textit{standalone case}.\\
\textbf{Cluster-Based Sampling (CBS):} For each annotated \textit{case}-excerpt pair, if the \textit{case} is part of a cluster with contrasting \textit{cases}, we first sample excerpts matched with other cases within the same cluster. We do this until we have either exhausted all possible unique negative samples\footnote {This can happen when only one of the \textit{cases} in a cluster has a large number of positive samples} or we have the required number of negative samples. When the number of negative samples is not satisfied, we sample from outside the cluster randomly. For standalone cases, we collect negative samples randomly 

We trained separate models for each sampling technique and sampling ratio. We experiment with sampling 1x, 2x, 3x, and 5x number of negative samples as positive samples. We divided the data into train, validation, and test sets in the ratio 3:1:1. We keep the test and validation sets constant across different sampling ratios, with 1:1 sampling. To train the model, we input the concatenated case \(C_i\) and segment \(S_j\) separated by a special token to fine-tune PrivBERT, a privacy policy language model \citep{srinath2021privacy}. We use a binary cross-entropy loss function to optimize the parameters and train the model using Adam optimizer \citep{kingma2014adam} with a learning rate of 1e-5 for 3 epochs. 

We evaluate the trained model on test sets each containing policy excerpts previously unseen by the model. The standalone set and the contrasting set contain case-excerpt pairs from standalone cases contrasting cases respectively. We compare the performance of our model to several approaches listed in Table \ref{baseline-eval}. 1) We use Sentence-BERT \citep{reimers2019sentence} to vectorize policy excerpts and ToS;DR cases. We then calculate the cosine distance between each pair and identify a threshold similarity score (0.25) based on the validation set. 2) We prompt OpenAI's GPT-4 to identify matching case-excerpt pairs 3) We enhance the GPT-4 prompt with two examples, one positive/negative case-excerpt pair 4) We train RoBERTa \citep{liu2019roberta} recreating the experiments with PrivBERT and report the results on the best performing model. 

\section{Results}

The results of PrivBERT trained on datasets created by random sampling (RS) and cluster-based sampling (CBS), for a 1:3 sampling ratio, is shown in Table \ref{held_out_eval}. The results in the table are reported after taking the mean (max standard deviation: 0.05) on 3 runs with different random initial states and training sets shuffled. The 0 category refers to disparate case-excerpt pairs while 1 refers to matching pairs. The table shows that the results on the standalone set are superior to those on the contrasting case set across both sampling techniques. This is likely since \textit{contrasting cases} are often lexically and semantically quite similar, and therefore a harder problem. We also see that cluster-based sampling performs significantly better than random sampling on the contrasting set while maintaining a similar performance on the standalone set. This is expected since a larger number of samples are used to train the model to solve the harder problem of disambiguating between similar cases. 

For the RS model, the results on the contrasting set improved with larger negative sampling ratios while the CBS model remained the same. This could indicate that a larger number of contrasting case training samples were necessary for the RS model. On the other hand, the results for both the models on the standalone set remained constant across different sampling ratios. Results for different sampling ratios are presented in the appendix. 

\begin{table}[ht]
\centering
\scalebox{1}{
\begin{tabular}{|c|c|c|c|c|}
\hline
\textbf{Model} & \textbf{C} & \textbf{P} & \textbf{R} & \textbf{F1}\\
\hline
Sentence-BERT & 0 & 0.61 & 0.66 & 0.63\\
& 1 & 0.78 & 0.71 & 0.74 \\
\hline
GPT-4 (Zero-shot) & 0 & 0.72 &  0.74 & 0.73\\
& 1 & 0.82 & 0.80 & 0.81\\
\hline
GPT-4 (2-shot) & 0 & 0.79 & 0.74 & 0.76\\
& 1 & 0.83 & 0.86 & 0.84\\
\hline
RoBERTa (CS) & 0 & 0.78 &  0.85 & 0.81\\
& 1 & 0.91 & 0.87 & 0.89\\
\hline
PrivBERT (CS) & 0 & \textbf{0.80} & \textbf{0.88} & \textbf{0.83}\\
& 1 & \textbf{0.95} & \textbf{0.89} & \textbf{0.92}\\
\hline
\end{tabular}}
\caption{\label{baseline-eval} Test performance (P: Precision, R: Recall)}
\end{table}

We achieve state-of-the-art results on this dataset, the performance of several other approaches is shown in Table \ref{baseline-eval}. Finetuned RoBERTa-base is the second best-performing model, followed by the two-shot version of GPT-4. The performance improvement due to fine-tuning depicts the complexity of the task. Sentence-BERT performs well on case-excerpt pairs with significant lexical overlap.

Under real-world conditions, identifying all data practices within a privacy policy would involve testing whether an excerpt matches any case in the full set of ToS;DR cases. It is therefore important that the model be trained robustly on dissimilar pairs, supporting training on larger sampling ratios and providing high precision scores for negatives. Furthermore, the erroneous classification of a pair as a `match' can be readily discerned and flagged, thus facilitating the establishment of a feedback loop. Conversely, the misclassification of a pair as dissimilar may evade detection, reinforcing our argument for the necessity of robustly performing models on dissimilar pairs.

\subsection{Case Study}

\begin{figure}[h]
    \centering
    \includegraphics[scale=0.34]{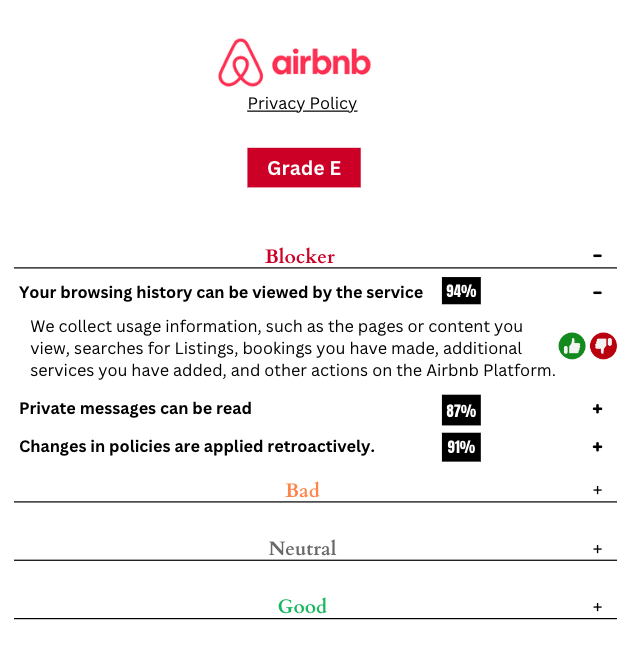}
    \caption{Privacy label with varying information levels. Detailed information access by the expand/contract feature. Model probabilities shown as \%; thumbs-up/down icons for user feedback on matches.
    }
    \label{fig:tool}
\end{figure}

We present a case study evaluating our model's performance on a previously un-annotated policy. We chose Airbnb\footnote{https://www.airbnb.com/}, a popular online marketplace, due to its popularity and its legitimate need to access sensitive personal user information. We split the policy based on new line characters, resulting in 150 segments encompassing paragraphs, section titles, and list items. Two authors of the paper thoroughly reviewed each segment and tagged them with any associated ToS;DR \textit{cases}. Subsequently, we employed the best-performing CBS model to identify matching \textit{case}-excerpt pairs. Table \ref{case_study} that the model consistently identified true positives, while producing a few false positives, therefore leading to a high recall for matching case-excerpt pairs. Most false positives were caused due to inaccurate predictions on privacy policy section titles. The model was able to identify true negatives with a high degree of accuracy while producing a few false negatives, leading to high precision scores for dissimilar case-excerpt pairs. 

\begin{table}[ht]
\centering
\scalebox{0.9}{
\begin{tabular}{|c|c|c|c|c|}
\hline
& \textbf{Precision} & \textbf{Recall} & \textbf{F1} & \textbf{Support} \\
\hline
\textbf{True} & 0.79 & 0.88 & 0.83 & 285 \\
\textbf{False} & 0.87 & 0.84 & 0.85 & 19,215 \\
\hline
\end{tabular}}
\caption{\label{case_study} Case-level results}
\end{table}

In Fig. \ref{fig:tool}, we demonstrate the tool that provides users with privacy information at varying levels of granularity. At the highest level, identified \textit{cases} are summarized using a letter grade with the grading scheme adopted by ToS;DR: 1 or more `blockers': Grade E || > 75\%  `bad' \textit{cases}: Grade D || 50\% - 75\% `bad' cases: Grade C || < 50\% `bad' cases: Grade B || < 25\% `bad' cases: Grade A. At the next level, identified \textit{cases} are listed under their rating categories. Finally, at the most fine-grained level policy excerpts associated with a case are listed along with a match probability score. 

\section{Conclusion}

Our research introduces a novel solution to the challenge of simplifying complex privacy policies. We provide an automated approach to scale identification of data practices in privacy policies, providing users with information regarding their data privacy. 
By automating the process of aligning data practices with policy excerpts, we are taking a crucial step towards eliminating the need for users to laboriously parse through lengthy and intricate documents, ensuring that they receive precise and easily digestible information about their online privacy. This lowers the barrier for internet users to understand what is happening with their data online, thereby allowing them to make informed privacy decisions. While there is room for further refinements and research, our model presents a promising foundation for future advancements in the field of usable privacy, ultimately promoting greater user empowerment and security in the digital age.

\section{Limitations}

Here, we attempt to match ToS;DR cases to their corresponding privacy policy segments. While the dataset we create attempts to simulate real-world conditions through the hold-out set, we match segments that are preciously identified by annotators. It is therefore possible that there is some loss of performance in a real-world setting. We rectify this issue in our case study by applying our model on automatically segmented policies and find that the performance is still comparable with that of the hold-out set. However, it is possible that the results in the case study might not scale over all privacy policies.

We use the rating scheme developed by ToS;DR which could potentially be biased. There is therefore a need for further research to understand user privacy preferences towards threshold for blocker, bad, good and neutral data practices. This extends to the assignment of letter grades. 

\section{Ethical Statement}

We do not foresee any ethical issues with our work.

\bibliography{latex/acl_latex}

\appendix

\section{Appendix}
\label{sec:appendix}

\subsection{Loss Function}

Given a \textit{case} from ToS;DR, \(C_i\), and an excerpt from a privacy policy, \(S_j\),  where,\\ \\
\(C\) \(\in\) {set of ToS;DR \textit{cases}}; \\
\(S\) \(\in\) {set of privacy policy excerpts} \\ \\
The loss function is given by, \\
\begin{equation}
    -{y\log(\hat{y}) + (1 - y)\log(1 - \hat{y})}
\end{equation}

\begin{equation}
    y = Match(C_i, S_j)
\end{equation}
\begin{table*}[h]
\centering
\scalebox{0.9}{
\begin{tabular}{|c|c|c|c|c|c|c|c|c|c|c|}
\hline
\multirow{2}{*}{\textbf{Ratio}} & \multirow{2}{*}{\textbf{Model}} & \multirow{2}{*}{\textbf{C}} &  \multicolumn{4}{|c|}{\textbf{Standalone Set}} & \multicolumn{4}{|c|}{\textbf{Contrasting Set}} \\
\cline{4-11}
 & & & \textbf{P} & \textbf{R} & \textbf{F1} & \textbf{S} & \textbf{P} & \textbf{R} & \textbf{F1} & \textbf{S} \\ 
\hline
\multirow{4}{*}{1:1}& \multirow{2}{*}{\(RS\)} & 0 & 0.94 & 0.93 & 0.93 & 2091 & 0.72 & 0.28 & 0.40 & 557\\
&                         & 1 & 0.93 & 0.94 & 0.94 & 2091 & 0.70 & 0.93 & 0.80 & 996\\
\cline{2-11}
& \multirow{2}{*}{\(CBS\)}& 0 & 0.93 & 0.93 & 0.93 & 2091 & 0.77 & 0.60 & 0.68 & 557\\
&                        & 1 & 0.93 & 0.92 & 0.92 & 2091 & 0.90 & 0.95 & 0.92 & 996\\
\hline
\multirow{4}{*}{1:2} & \multirow{2}{*}{\(RS\)} & 0 & 0.95 & 0.95 & 0.95 & 2091 & 0.75 & 0.30 & 0.43 & 557\\
&                         & 1 & 0.95 & 0.95 & 0.95 & 2091 & 0.70 & 0.94 & 0.81 & 996\\
\cline{2-11}
& \multirow{2}{*}{\(CBS\)}& 0 & 0.87 & 0.94 & 0.90 & 2091 & 0.70 & 0.81 & 0.75 & 557\\
&                        & 1 & 0.94 & 0.86 & 0.90 & 2091 & 0.94 & 0.91 & 0.92 & 996\\
\hline

\multirow{4}{*}{1:3} & \multirow{2}{*}{\(RS\)} & 0 & 0.94 & 0.95 & 0.94 & 2091 & 0.75 & 0.37 & 0.50 & 557\\
&                         & 1 & 0.95 & 0.94 & 0.94 & 2091 & 0.72 & 0.93 & 0.81 & 996\\
\cline{2-11}
& \multirow{2}{*}{\(CBS\)}& 0 & 0.89 & 0.95 & 0.92 & 2091 & 0.71 & 0.82 & 0.76 & 557\\
&                        & 1 & 0.95 & 0.88 & 0.91 & 2091 & 0.95 & 0.91 & 0.93 & 996\\
\hline

\multirow{4}{*}{1:5} & \multirow{2}{*}{\(RS\)} & 0 & 0.95 & 0.95 & 0.95 & 2091 & 0.69 & 0.45 & 0.54 & 557\\
&                         & 1 & 0.93 & 0.94 & 0.93 & 2091 & 0.70 & 0.92 & 0.78 & 996\\
\cline{2-11}
& \multirow{2}{*}{\(CBS\)}& 0 & 0.86 & 0.96 & 0.90 & 2091 & 0.71 & 0.83 & 0.76 & 557\\
&                        & 1 & 0.92 & 0.88 & 0.89 & 2091 & 0.90 & 0.90 & 0.90 & 996\\
\hline
\end{tabular}}
\caption{\label{held_out_eval} Performance of PrivBERT on random sampling and cluster-based sampling over various sampling ratios (P: Precision, R: Recall, S: Support)}
\end{table*}

\subsection{Privacy Rating Analysis}

\begin{figure}[h]
    \centering
    \includegraphics[scale=0.28]{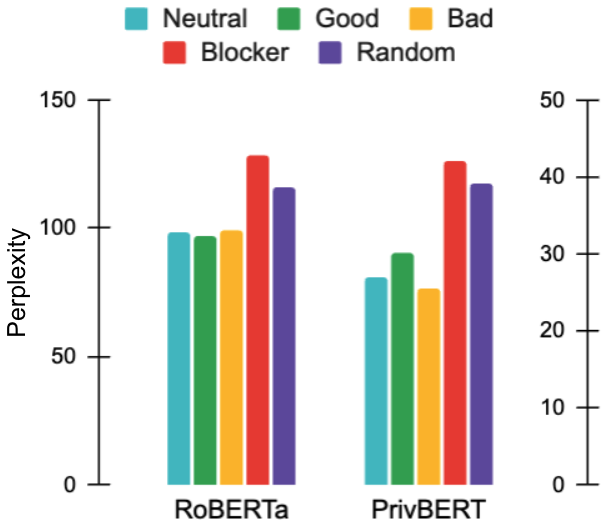}
    \caption{Perplexity distribution of ToS;DR ratings}
    \label{fig:perplexity}
\end{figure}

We further investigated each ToS;DR privacy category, blocker, bad, good, and neutral by calculating the perplexity of the excerpts associated with them. Figure \ref{fig:perplexity} shows the average perplexity of excerpts calculated using RoBERTa, a general purpose language model and PrivBERT, a language model pre-trained on privacy policy text. For comparison, we randomly sampled 1000 sentences from ten most visited websites' policies. We calculated the perplexity by masking each word in the sequence separately and exponentiating the average the negative log likelihood of size 8 sequences with a stride of 4.

From Figure \ref{fig:perplexity}, it is evident that PrivBERT exhibits notably reduced perplexity scores in comparison to RoBERTa, a phenomenon attributed to PrivBERT's pretraining on policy text. We hypothesize that for PrivBERT (but not RoBERTa), text segments commonly encountered in privacy policies should yield lower perplexity scores, while those that are infrequent should result in higher perplexity scores. Figure \ref{fig:perplexity} shows that `blockers' have the highest sentence perplexity scores for both language models suggesting that they might be written in particularly convoluted language unusual even within privacy policies. Conversely, other categories exhibit closely aligned perplexity scores for RoBERTa, but distinct variations for PrivBERT, suggesting that the pretraining data may encompass text associated with each category at varying frequencies. Notably, `bad' practices are the second most perplexing category for RoBERTa, yet the least perplexing for PrivBERT, signifying the frequent inclusion of `bad' practices-related text in policies. Conversely, this trend is reversed for `good' practices, indicating their relative rarity.


\subsection{Prompts Used for GPT-4}

We used the OpenAI API GPT-4 models with a temperature setting of 0. \\

\noindent\fbox{%
    \parbox{\textwidth}{%

\textcolor{gray}{You are a privacy policy expert. Given a title and quote, your task is to evaluate whether the title represents the data practice described in the quote. Your output should only be either 0 (indicating the title does not represent the quote) or 1 (indicating the title represents the quote).}\\

Example 1\\
\textcolor{gray}{Title: Third parties are involved in operating the service. \\Quote: Note that we don't use any 3rd party website statistics tools like Google Analytics or similar.}\\
Output: \textcolor{gray}{1} \\

Example 2\\
\textcolor{gray}{Title: You can opt out of targeted advertising\\
Quote: Our CDN is Cloudflare, and they may include cookies with our pages to provide a better service.}\\
Output: \textcolor{gray}{0}\\

\textcolor{gray}{Title: <title>\\
Quote: <quote>}

}}

\end{document}